# A Low-throughput Wavelet-based Steganography Audio Scheme

P. Carrión[1], H.M. de Oliveira[1], R.M. Campello de Souza[1]

[1]Federal University of Pernambuco - UFPE, C.P. 7.800, 50.711-970, Recife - PE, Brazil

jpcc1@hotmail.com, {hmo, ricardo}@ufpe.br

## 1. Waiter, please: There is a text in my audio

Steganography literally means "covered writing", which is derived from the Greek στεγοσ (roof) and γραπηια (writing). Cryptography, watermarking and stenography are closely related [Johnson and Jajodia 1998]. This paper introduces the basis of a wavelet-based low-throughput secret key steganography system that requires the exchange of a secret key (stego-key) prior to communication. We try to take apart the fact that the message can only be read with a secret key by using a standard cryptosystem [Schneier 1996] before hiding a text in an audio file. An interesting recent survey lists about 32 different audio stego-tools (.wav 50%, .mp3 28%, midi 6%, others 16%) [Dittmann and Kraetzer 2006]. Possible strategies for embedding data inside the host audio include the use of frequencies inaudible to humans [Gopalan and Wenndt 2004], embedding data using the LSB [Dumitrescu et al. 2003], transform embedding techniques [Lin and Delp 1999]. This paper presents the preliminary of a novel scheme of steganography, and introduces the idea of combining two secret keys in the operation (Figure 1). The first secret key encrypts the text using a standard cryptographic scheme (e.g. IDEA, SAFER+, etc.) prior to the wavelet audio decomposition [de Oliveira 2007]. The way in which the cipher text is embedded in the file requires another key, namely a stego-key, which is associated with features of the audio wavelet analysis.

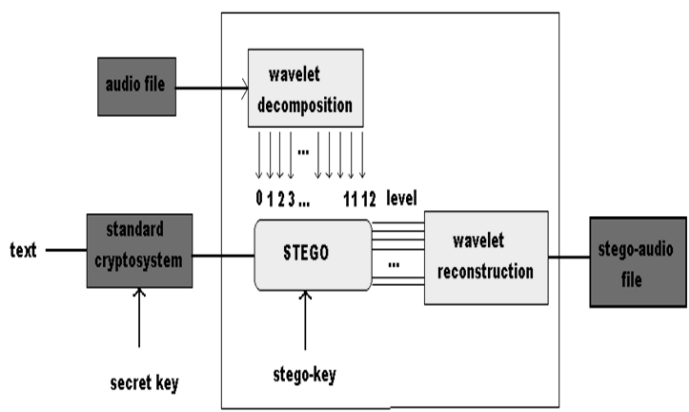

**Figure 1. A method of embedding cover text in an audio file by wavelet transform**

## 2. On implementing the Scheme

A program was written using the Matlab[TM] platform [Kamen and Heck 1997] for embedding secret information into a sound, which is carried out by wavelet analysis. Two matrices are generated in the audio decomposition– the first one ($n \times 1$) contains wavelet coefficients, the column specifies the number of channels (mono, in this prospective study), and the second furnishes the length of decomposed signals in each level. Differently from previous schemes [Dittmann and Kraetzer 2006, Noto 2008], data are used to **replace** the three decimal positions of the wavelet coefficients of the audio decomposition matrix. The text is converted into a decimal string using the ASCII code. It is then parsed and each subblock allows allocating information at different wavelet scales. Different alphanumeric passwords are provided to each decomposition level so as to spread the hidden information. Longer passwords are required at wavelet decomposition levels where much information is selected to be embedded. Another part of the stego-key gives the "start position" where information will be placed at each selected decomposition level. Replacing the first three decimal positions in the floating-point representation of the wavelet coefficients carries out data insertion. A similar procedure is performed at the other levels. Both the source code (extension .m) developed to implement this steganography technique and a straightforward example are freeware available at the URL http://www2.ee.ufpe.br/codec/stego.html. In this version, only the stego-key is provided and the secret key should be supplied by external available software.

## Acknowledgements

The authors are grateful to Mr. Jorge Rehn. CNPq research grant #301996 (HMdO).